\documentclass[twocolumn,showpacs,english,superscriptaddress,prl,preprintnumbers,amsmath,amssymb,floatfix]{revtex4-1}

\usepackage[T1]{fontenc}
\usepackage[latin9]{inputenc}
\usepackage{dcolumn}
\usepackage{bm}
\usepackage{graphicx}
\usepackage{color}
\usepackage{esint}
\usepackage{babel}
\usepackage{amsfonts}
\usepackage{slashed}
\usepackage{enumerate}

\begin{document}

\title{\bm{$6\pi$} Josephson effect in Majorana box devices}

\author{A.~Zazunov}
\affiliation{Institut f\"ur Theoretische Physik,
Heinrich-Heine-Universit\"at, D-40225  D\"usseldorf, Germany}
\author{F. Buccheri}
\affiliation{Institut f\"ur Theoretische Physik,
Heinrich-Heine-Universit\"at, D-40225  D\"usseldorf, Germany}
\author{P.~Sodano}
\affiliation{International Institute of Physics, Universidade Federal do Rio Grande do
Norte, 59012-970 Natal, Brazil}
\affiliation{INFN, Sezione di Perugia, Via A.~Pascoli, I-06123 Perugia, Italy}
\author{R.~Egger}
\affiliation{Institut f\"ur Theoretische Physik,
Heinrich-Heine-Universit\"at, D-40225  D\"usseldorf, Germany}
\date{\today}

\begin{abstract}
We study Majorana devices featuring a competition between superconductivity and multi-channel Kondo physics. Our proposal extends previous work on single-channel Kondo systems to a topologically nontrivial setting of non-Fermi liquid type, where topological superconductor wires (with gap $\Delta$) represent leads tunnel-coupled to
a Coulomb-blockaded Majorana box.  On the box, a spin degree of freedom
with Kondo temperature $T_K$ is nonlocally defined in terms of Majorana states. For $\Delta\gg T_K$, the destruction of
Kondo screening by superconductivity implies a $4\pi$-periodic Josephson current-phase relation. Using a strong-coupling analysis  in the opposite regime
 $\Delta \ll T_K$, we find a $6\pi$-periodic Josephson relation for three leads, with critical current $I_c\approx e\Delta^2/ \hbar T_K$,
 corresponding to the transfer of fractionalized charges $e^*=2e/3$.
\end{abstract}

\pacs{74.50.+r, 74.78.Na, 72.15.Qm, 75.20.Hr}

\maketitle

\emph{Introduction.---}An important goal of condensed matter physics and quantum information science is to implement, thoroughly understand, and usefully employ
systems hosting topologically protected Majorana bound states (MBSs)  \cite{Alicea2012,Leijnse2012,Beenakker2013}. These states are  expected
  near the ends of topological superconductor (TS) wires, and
experimental evidence for MBSs has been reported for semiconductor-superconductor heterostructures with proximitized InAs or InSb nanowires
 \cite{Mourik2012,Krogstrup2015,Higginbotham2015,Albrecht2016,Leo2016}.
For a Coulomb-blockaded superconducting island containing more than two MBSs (`Majorana box'), a spin operator is encoded
by pairs of spatially separated MBSs.  When normal leads are coupled to the MBSs, this spin is screened through cotunneling processes,
culminating in the so-called topological Kondo effect (TKE) \cite{Beri2012,Altland2013,Beri2013,Crampe2013,Tsvelik2013,Zazunov2014,Altland2014,
Eriksson2014,Galpin2014,Buccheri2015,Buccheri2016,Giuliano2016a,Giuliano2016b}
which exhibits non-Fermi liquid physics below $T_K$.  Unlike other overscreened multi-channel Kondo systems \cite{Gogolin1998,Potok2007,Pierre2015,Keller2015}, the TKE
 is intrinsically stable against anisotropies.
Majorana devices could thus realize multi-channel Kondo effects without delicate fine tuning of parameters.

\begin{figure}[t]
\centering
\includegraphics[width=\columnwidth]{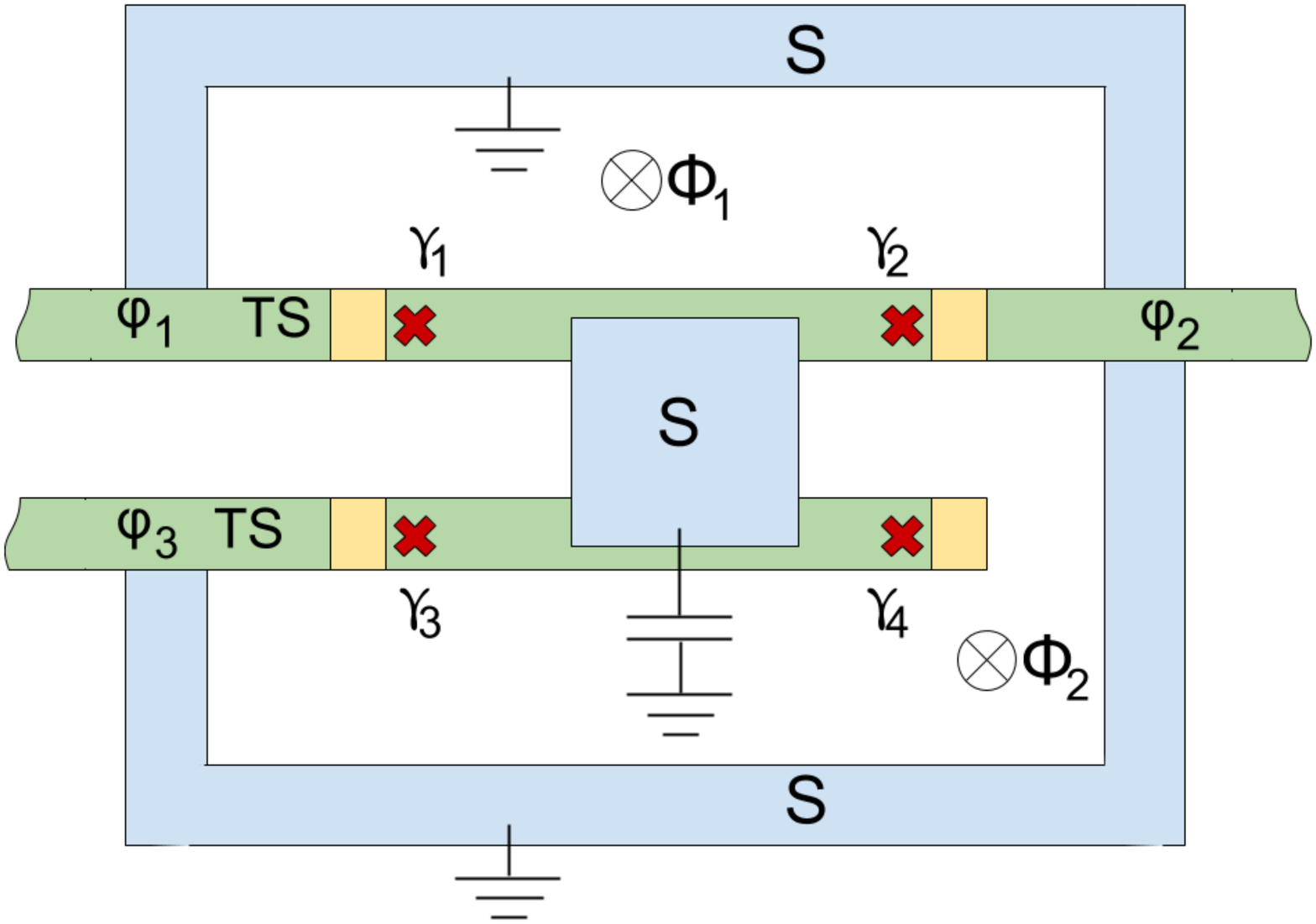}
\caption{Schematic device setup with $M=3$ superconducting leads,
using two long parallel InAs (or InSb) nanowires.
Proximitized parts give TS wire regions (green) with Majorana end states (red crosses,
shown only on the parts forming the Majorana box). Short non-proximitized sections (yellow) are used as gate-tunable tunnel contacts. The floating Majorana box with
four MBSs ($\gamma_j$) is created by joining the two central TS parts through an $s$-wave superconducting bridge (blue).
 Superconducting leads are obtained from outer TS wire sections in contact with
conventional superconductors (blue). By tuning magnetic fluxes ($\Phi_{1,2}$),
the supercurrents $I_j$ can be studied as function of the
phases $(\varphi_1,\varphi_2,\varphi_3)$.}
\label{f1}
\end{figure}

We here study the Josephson effect for a Majorana box with
superconducting (instead of normal) leads as illustrated in Fig.~\ref{f1}.
Previous theoretical work for Majorana systems contacted by superconducting electrodes has only addressed cases without TKE~\cite{Zazunov2012,Peng2015,Ioselevich2016,Zazunov2016,Setiawan2016}.
In our setup, a nontrivial competition between superconductivity and the Kondo effect
arises because lead states below the superconducting gap $\Delta$ are not
available anymore for screening the box spin.  The simpler single-channel spin-$1/2$ Kondo case, which
 is of Fermi-liquid type and can be realized when two superconducting leads are connected to a quantum dot \cite{Alvaro2011},
 was studied in detail both theoretically \cite{Glazman1989,Golub1996,Rozhkov1999,Vecino2003,Siano2004,Choi2004,Karrasch2008,Luitz2012} and experimentally \cite{Kasumov1999,VanDam2006,Cleuziou2006,Jorgensen2007,Eichler2009,Delagrange2015}.
It has been established that a local quantum phase transition at $\Delta/T_K\simeq 1$ separates a
so-called 0-phase (small $\Delta/T_K$) and a $\pi$-phase (large $\Delta/T_K$), where essentially the entire crossover is
 described by universal scaling functions of $\Delta/T_K$. Deep in the 0-phase, the Kondo resonance persists and yields
 the current-phase relation of a fully transparent superconducting junction,
 while in the $\pi$-phase the Kondo effect is almost completely quenched and one finds a negative supercurrent.

With the Majorana device proposed below, the rich interplay between superconductivity and multi-channel Kondo screening
may also become experimentally accessible. The symmetry group of the
TKE is here affected by even a tiny gap $\Delta$ due to the proliferation of crossed Andreev reflection processes.
  For $\Delta\ll T_K$ and $M=3$ attached leads, our nonperturbative strong-coupling theory predicts that two-channel Kondo physics is responsible for
  a $6\pi$-periodic Josephson effect  with critical current $I_c\approx e\Delta^2/\hbar T_K$.
 This periodicity implies charge fractionalization  in units of $e^*=2e/3$ for  elementary transfer processes.
On the other hand, for $\Delta\gg T_K$, we recover the well-known
 $4\pi$-periodic current-phase relation of parity-conserving
 topological Josephson junctions \cite{Alicea2012,Leijnse2012,Beenakker2013}.
In view of the rapid experimental progress on Majorana states in semiconductor-superconductor devices
\cite{Mourik2012,Krogstrup2015,Higginbotham2015,Albrecht2016,Leo2016}, our predictions can likely be tested soon, e.g., by the
techniques recently employed to observe the $4\pi$ Josephson effect \cite{Bocquillon2016}.

\emph{Model.---}The superconducting leads attached to the Majorana box are
described as semi-infinite TS wires of symmetry class $D$.
 For $M$ leads, the effectively spinless
low-energy Hamiltonian is (we put $e=\hbar=v_F=1$) \cite{Alicea2012}
\begin{equation}\label{leadsHam}
H_{\rm leads}=\sum_{j=1}^M\int_0^\infty dx  \Psi^\dagger_j (x) \left[-i \partial_x \sigma_z
+ \Delta_j e^{-i\varphi_j\sigma_z} \sigma_y\right] \Psi^{}_j(x) ,
\end{equation}
where $\Delta_j$ denotes the absolute value and $\varphi_j$ the phase of the respective proximity-induced superconducting gap,  Pauli matrices $\sigma_{x,y,z}$ and unity $\sigma_0$ act in Nambu space, and the spinors $\Psi_j=(\psi^{}_{j,R},\psi_{j,L}^\dagger)^T$ are expressed in terms of right/left-moving fermion operators with boundary condition $\psi_{j,R}(0)=\psi_{j,L}(0)$.  We mainly discuss results for
identical gaps, $\Delta_j=\Delta$, but our theory applies to the general case
\cite{SM1}.
The reason why we did not assume conventional $s$-wave superconductors as leads is that different pairing symmetries for the box and the leads imply a supercurrent blockade \cite{Zazunov2012}, where only above-gap quasiparticle transport is possible under rather general conditions \cite{Peng2015,Ioselevich2016,Zazunov2016,Setiawan2016}.
   Fortunately, leads with effective  $p$-wave pairing symmetry may be implemented in a natural way, see below and Fig.~\ref{f1}.
At $x=0$, each lead fermion $\psi_j$ is then coupled by a tunnel amplitude $t_j$
to the respective Majorana operator $\gamma_j=\gamma_j^\dagger$ on the box, with anticommutator $\{\gamma_j,\gamma_k\}=\delta_{jk}$.
We study energy scales well below the proximity gap $\Delta_{\rm box}$ on the box, where $\Delta_{\rm box}$ and $\Delta\alt \Delta_{\rm box}$ are taken as independent parameters and above-gap quasiparticles on the box are neglected.
For a large charging energy $E_C$, charge quantization
implies a parity constraint for the Majorana states
on the box and tends to suppress quasiparticle poisoning processes.
Nonetheless, the ground state remains degenerate for $M>2$, where the Majorana bilinears $i\gamma_j\gamma_k$ represent
the box spin \cite{Beri2012,Altland2014}.
The projection to the Hilbert space sector with quantized box charge yields \cite{Beri2012}
 \begin{equation}\label{ham1}
 H_{\rm EC}=\sum_{j\ne k}^M \lambda_{jk} \psi^\dagger_j(0) \psi^{}_k(0)\gamma_k\gamma_j,
\end{equation}
where the dimensionless exchange couplings $\lambda_{jk}= 2t_j t^\ast_k/E_C$ describe elastic cotunneling between leads $j\leftrightarrow k$. For $\Delta=0$, Refs.~\cite{Beri2012,Altland2013,Beri2013} show that
$H_{\rm leads}+H_{\rm EC}$ gives a TKE of
 SO$_2(M)$ symmetry with
\begin{equation}\label{tkdef}
T_K= E_C e^{ -\pi/[2(M-2) \bar \lambda]},\quad \bar\lambda=\frac{1}{M(M-1)}\sum_{j\ne k} \lambda_{jk},
\end{equation}
where the group relation SO$_2(3)\sim$~SU$_4(2)$ implies a four-channel Kondo effect for $M=3$ \cite{Fabrizio1994}.  For $\Delta\ne 0$, the competition between Kondo physics and superconductivity is then controlled by the ratio $\Delta/T_K$.  For $\varphi_j=0$,
the above model also describes junctions of off-critical anisotropic spin chains \cite{Crampe2013,Tsvelik2013,Giuliano2016a,Giuliano2016b}.

\emph{Implementation.---}Before turning to results, we briefly discuss
how to realize this model for the simplest nontrivial case $M=3$,
cf.~Fig.~\ref{f1}.  The floating box is defined by connecting two
parallel TS wires by an $s$-wave superconductor. Nanowires can be fabricated
with an epitaxial superconducting shell \cite{Krogstrup2015}, where a
magnetic field simultaneously drives both wires into the TS phase \cite{Alicea2012}. We assume that the TS sections on the box are so
long that overlap between different Majorana states is negligible. Non-proximitized wire parts yield gate-tunable tunnel barriers, and leads are defined by the outer TS wires in Fig.~\ref{f1}. Using available Majorana wires \cite{Albrecht2016}, it appears possible to
realize the Kondo regime \cite{Beri2012,Zazunov2014,Altland2014,Beri2016}.
 In a loop geometry with magnetic fluxes \cite{Giazotto}, one can change the phase differences between TS leads and measure the current-phase relation.

\emph{Josephson current.---}It is often convenient to integrate out the lead fermion modes away from $x=0$.  The Euclidean action, $S=S_{\rm leads}+S_{\rm box}+S_{\rm EC}$, is thereby expressed in terms of Majorana
fields $\gamma_j(\tau)$ and boundary ($x=0$) Grassmann-Nambu spinor fields,
$\Psi_j(\tau)=\left(\psi_j,\bar\psi_j\right)^T$.
With inverse temperature $\beta$, $H_{\rm leads}$ gives
\begin{equation}  \label{sleads}
S_{\rm leads}= -\frac12 \int_0^\beta d\tau d\tau' \sum_j \bar \Psi_j(\tau) G^{-1}_j(\tau-\tau') \Psi_j(\tau') ,
\end{equation}
where the boundary Green's function $G_j(\tau)$ has the  Fourier transform \cite{Zazunov2016}
\begin{equation} \label{bGF}
G_j(\omega) = -i\ {\rm sgn}(\omega) \sqrt{1+\frac{\Delta_j^2}{\omega^2}} \ \sigma_0
+\frac{\Delta_j e^{-i\varphi_j\sigma_z}}{i\omega} \sigma_x .
\end{equation}
 The box action is $S_{\rm box}=\frac12\int d\tau \sum_j \gamma_j\partial_\tau
\gamma_j$, and $S_{\rm EC}=\int d\tau H_{\rm EC}(\tau)$.
Expressing the partition function as functional integral,
$Z= e^{-\beta F} = \int {\cal D}(\Psi_j,\gamma_j) e^{-S}$,
the supercurrent $I_j$ through lead no.~$j$ (oriented towards the box) follows as
phase derivative of the free energy, $I_j=(2e/\hbar) \partial_{\varphi_j} F,$
where current conservation implies $\sum_j I_j=0$.
We discuss the zero-temperature limit in what follows.

\emph{Atomic limit: $\Delta\gg T_K$.---}In the atomic limit,
 after gauging out the $\varphi_j$ phases from the bulk, Eq.~(\ref{bGF})
simplifies to $G_j(\omega) \simeq \frac{\Delta_j}{i\omega}(\sigma_0+\sigma_x)$.
  As a consequence,  the lead action \eqref{sleads} becomes
$S_{\rm leads}\simeq \frac12\sum_j \int d\tau \ \eta_j\partial_\tau\eta_j$,
where the boundary fermions define zero-energy
Majorana operators, $\eta_j= (\psi_j+\psi_j^\dagger)/\sqrt{2\Delta_j}$, and Eq.~\eqref{ham1} yields the effective low-energy Hamiltonian
$H_{\rm eff}=\frac{1}{2}  \sum_{j\ne k} \sqrt{\Delta_j\Delta_k} \lambda_{jk} e^{i(\varphi_j-\varphi_k)/2} \eta_j\eta_k\gamma_k\gamma_j.$
Since all mutually commuting products $2i\eta_j\gamma_j=\sigma_j=\pm 1$ are conserved, we arrive at the $4\pi$-periodic supercurrents
\begin{equation}\label{4pi}
I_j=\frac{e\Delta}{\hbar}\sum_{k\ne j}\lambda_{jk} \sigma_j\sigma_k\sin\left( \frac{\varphi_j-\varphi_k}{2}\right),
\end{equation}
where the $\{\sigma_j\}$ correspond to different fermion parity sectors and we put $\Delta_j=\Delta$.
The Kondo effect is suppressed in the atomic limit since no low-energy quasiparticles in the leads  are available to screen the box spin.
In fact, Eq.~\eqref{4pi} also describes topological Josephson junctions with
featureless tunnel contacts \cite{Alicea2012}.

\emph{Renormalization group (RG) analysis.---}To tackle the case of arbitrary $\Delta/T_K$, we start with the one-loop RG equations.
Renormalizations appear for $\Delta$, for the $\lambda_{jk}$,
and for the complex-valued crossed Andreev reflection amplitudes $\kappa_{jk}=\kappa_{kj}$ (with $j\ne k$). Such couplings are
absent in the bare model but will be generated during the RG flow by an interplay of
exchange processes ($\lambda$) and superconductivity ($\Delta$).
They describe the creation (or annihilation) of two fermions in
different leads by splitting (or forming) a Cooper pair on another lead,
corresponding to the additional term
\begin{equation}\label{carH}
H_{\rm CAR} = \sum_{j<k} \kappa_{jk} \psi_j^\dagger(0) \psi_k^\dagger(0)\gamma_k\gamma_j + {\rm h.c.}
\end{equation}
 In the Supplementary Material \cite{SM1}, we provide a derivation of the RG equations for arbitrary $\Delta_j\ll D$, where the bandwidth is $D\simeq {\rm min}(E_C,\Delta_{\rm box})$. For $\Delta_j=\Delta$, taking a gauge where the phase dependence only appears in $H_{\rm EC}+H_{\rm CAR}$, we  obtain ($j\ne k$)
\begin{eqnarray} \nonumber
\frac{d \lambda_{jk}}{d \ell} &=& \frac{2}{\pi} \sum_{m\ne(j,k)}^M\Bigl [
\sqrt{1+\delta^2}\left ( \lambda_{jm} \lambda_{mk} +  \kappa_{jm} \kappa_{mk}^\ast\right)\\
& +& \label{rgeq2}
\delta \left(\lambda_{jm} \kappa_{mk}^\ast +  \kappa_{jm} \lambda_{mk}
\right)\Bigr],\\
  \nonumber
\frac{d \kappa_{jk} }{ d \ell} &=&\frac{2}{\pi}\sum_{m\ne (j,k)}^M\Bigl [
 \delta\left(\lambda_{jm} \lambda_{mk}^\ast + \kappa_{jm} \kappa_{mk} \right)\\
 \nonumber &+& \sqrt{1+\delta^2}
\left(\lambda_{jm} \kappa_{mk} + \kappa_{jm} \lambda_{mk}^\ast\right)\Bigr],
\end{eqnarray}
with $\delta(\ell)=e^\ell \Delta/D$ and initial conditions $\kappa_{jk}(0)=0$ and
$\lambda_{jk}(0)= (2t_jt^*_k/E_C) e^{i(\varphi_j-\varphi_k)/2}$. The RG flow thus only  depends on the gauge-invariant phase differences $\varphi_j-\varphi_k$.

\textit{RG solution for the unbiased case.---}Putting all $\varphi_j=0$, the above RG equations can be solved analytically.  The matrices $\Lambda_{jk}^{(\pm)} = \lambda_{jk} \pm \kappa_{jk}$ may now be chosen real symmetric and obey decoupled flow equations,
\begin{equation}\label{RG1}
\frac{d\Lambda^{(\pm)}_{jk}}{d\ell} =  \frac{2}{\pi}
(\sqrt{1+\delta^2}\pm \delta ) \sum_{m\ne (j,k)}^M \Lambda_{jm}^{(\pm)}\Lambda_{mk}^{(\pm)} ,
\end{equation}
which (up to a rescaling) coincide with those for the TKE. The results of Refs.~\cite{Beri2012,Altland2013,Beri2013,Zazunov2014}
imply
that anisotropies in the $\Lambda_{jk}^{(\pm)}$ are irrelevant perturbations, and both matrices scale towards isotropy,  $\Lambda^{(\pm)}_{jk}(\ell)\to \Lambda_\pm(\ell)[1-\delta_{jk}]$. For an isotropic initial condition, $\Lambda_\pm(0)=\bar \lambda$, with the average coupling $\bar \lambda$ in Eq.~\eqref{tkdef}, we find from Eq.~\eqref{RG1}
\begin{equation}\label{flowres}
\Lambda_\pm(\ell) = \lambda(\ell) \pm \kappa(\ell)= \frac{\bar\lambda}{1-\frac{2(M-2)\bar \lambda}{\pi} {\cal F}_\pm(\ell)},
\end{equation}
with the monotonically increasing functions
\begin{equation}
{\cal F}_\pm(\ell) = \left[ \sqrt{1+\delta^2}+ \ln
\sqrt{\frac{\sqrt{1+\delta^2}-1}{\sqrt{1+\delta^2}+1}}
 \pm \delta\right]^{\delta(\ell)}_{\delta(0)}.
\end{equation}
Hence $\Lambda_\pm(\ell)$ as well as $\delta(\ell)$ scale towards strong coupling
(with $\Lambda_+> \Lambda_-$).
For $\Delta\ll T_K$ with $T_K$ in Eq.~\eqref{tkdef},
the energy scales $T_\pm$ where $\Lambda_\pm(\ell)$ enters the strong-coupling regime can be estimated as
$T_\pm \simeq T_K e^{\pm \frac{\pi\Delta}{2(M-2)\bar \lambda E_C}}$.
The renormalized couplings $\lambda\agt \kappa$ are then of order unity
when reaching the strong-coupling regime. Now any finite coupling $\kappa$ (as well as $\Delta$) is expected to destabilize the SO$_2(M)$ Kondo fixed point and to induce a
flow to a stable fixed point with symmetry group SO$_1(M)$. For $M=3$, this has been shown in Ref.~\cite{Giuliano2016a}, where the relation SO$_1(3)\sim$~SU$_2(2)$ implies  a two-channel (instead of the $\Delta=0$ four-channel \cite{Beri2012,Fabrizio1994}) Kondo fixed point.
On the other hand, for $\Delta\gg T_K$, the pairing variable $\delta(\ell)$  reaches the strong-coupling regime first and we are back to the atomic limit.  In the remainder, we  discuss the limit $\Delta\ll T_K$ for $M=3$ leads.

\textit{Phase-biased case.---}For $\varphi_j\ne 0$, the
RG equations (\ref{rgeq2}) are more difficult to solve.
Numerical analysis of Eq.~\eqref{rgeq2} shows that for $\Delta\ll T_K$, the absolute values of the couplings $\lambda_{jk}$ and $\kappa_{jk}$ again flow towards isotropy but with a specific phase dependence.  With the real positive couplings $\lambda(\ell)$ and $\kappa(\ell)$ in Eq.~\eqref{flowres}, we find
$\lambda_{jk}(\ell) \to \lambda(\ell) e^{i(\varphi_j-\varphi_k)/2}$ and
$\kappa_{jk}(\ell)\to \kappa(\ell) e^{i\theta_{jk}}$ as one approaches the strong-coupling regime, where  $\theta_{jk}=(\varphi_j+\varphi_k)/2-\varphi_0$
with the center-of-mass phase $\varphi_0=(\varphi_1+\varphi_2+\varphi_3)/3$.
This result for $\theta_{jk}$ follows directly from gauge invariance and a
stationarity condition \cite{SM1}.
Finally, in what follows, it is convenient to remove the
phase factors from $H_{\rm EC}+H_{\rm CAR}$ by the gauge transformation
$\psi_{j,R/L}(x)\to e^{i(\varphi_j-\varphi_0)/2}\psi_{j,R/L}(x)$.

\emph{Strong-coupling analysis for $M=3$ and $\Delta\ll T_K$.---}We now turn to the asymptotic low-energy regime which can be accessed by perturbation theory
around the two-channel Kondo fixed point \cite{Eriksson2014,Affleck1993,Coleman1995}.
We first introduce chiral fermion fields for the TS leads by an unfolding transformation, $\Phi_j(x>0)=\psi_{j,R}(x)$ and $\Phi_j(x<0)=\psi_{j,L}(-x)$, and switch to their Majorana representations, $\Phi_j(x)=[\eta_j(x)+i\xi_j(x)]/\sqrt2$. Using the renormalized couplings $\Lambda_\pm=\lambda\pm\kappa$ in Eq.~\eqref{flowres} with
$\Lambda_+\gg  \Lambda_-$, we then obtain
\begin{equation}\label{hminus}
  H_{\rm EC}+H_{\rm CAR}=
  \Lambda_+ {\bf S}\cdot{\bf S}_\eta +\Lambda_- {\bf S}\cdot {\bf S}_\xi,
\end{equation}
where we define the spin-$1/2$ operators ${\bf S}=-\frac{i}{2}{\bm\gamma}\times
{\bm \gamma}$, ${\bf S}_\eta= - \frac{i}{2}{\bm \eta}(0)\times {\bm \eta}(0)$, and
 ${\bf S}_\xi=-\frac{i}{2}{\bm \xi}(0)\times {\bm \xi}(0)$, with
${\bm \gamma}=(\gamma_1,\gamma_2,\gamma_3)^T$ and similarly
 for the ${\bm \eta}$ and ${\bm \xi}$ Majorana triplets.
 The theory for $\Delta=\Lambda_-=0$ then describes the two-channel Kondo problem.
 The strong-coupling regime is accessible by employing
the following rules \cite{Eriksson2014,Affleck1993,Coleman1995}: (i)
Screening processes leading to a singlet state between ${\bm S}$ and ${\bm S}_\eta$ imply the replacement ${\bm S}\to iT_K^{-1/2}\gamma_0 {\bm \eta}(0)$, where the Majorana operator $\gamma_0$ describes the residual unscreened spin.
With time ordering ${\cal T}$, we have $\langle {\cal T}\gamma_0(\tau)
\gamma_0(0)\rangle=\frac12 {\rm sgn}(\tau)$.
   (ii) The Majorana triplet ${\bm \eta}(x)$ obeys twisted boundary
   conditions, ${\bm \eta}(x)\to {\rm sgn}(x) {\bm \eta}(x)$,
while the ${\bm \xi}$ triplet remains unchanged.  In terms of fermions, this implies perfect Andreev reflection, $\psi^{}_{j,R}(0)=-\psi^\dagger_{j,L}(0)$. (iii) For $\Delta=\Lambda_-=0$, the leading irrelevant operator is given by $H_{\rm LIO}=2\pi T_K^{-1/2} \gamma_0 \eta_1(0)\eta_2(0)\eta_3(0)$, with scaling dimension $d=3/2$.  The perturbation $H_-$ due to $\Lambda_-$, see Eq.~\eqref{hminus}, is then also irrelevant  with $d=3/2$.

We now have to include the bulk pairing term $\propto \Delta$ in the leads in a nonperturbative manner.  In fact, the leading contribution to $I_j$ follows from second-order perturbation theory in $H'=H_{\rm LIO}+H_{-}$. Since $H'$ has scaling dimension $d=3/2$, one naively expects a linear temperature $(T$) dependence of $I_j$.  However,  $\Delta$ is RG-relevant and provides a $1/T$ factor, resulting in a finite supercurrent at $T=0$.
We then need the boundary Green's functions for the field combinations $[\psi_{j,R}(0)-\psi_{j,L}^\dagger(0)]/\sqrt2$ representing decoupled TS leads with twisted boundary conditions. Following the steps in Ref.~\cite{Zazunov2016},
we thereby obtain the lead Majorana correlation functions at the boundary ($x=0^+$),
\begin{eqnarray}\label{correlationfunc}
&&\langle {\cal T}\eta_j(\tau)\xi_k(0)\rangle = -i
\delta_{jk} \Delta\cos(\varphi_j-\varphi_0) f(\tau),\\
 \nonumber
&& \langle {\cal T}\eta_j(\tau)\eta_k(0)\rangle =
 \langle {\cal T}\xi_j(\tau)\xi_k(0)\rangle = - \delta_{jk}  \partial_\tau f(\tau) ,
\\ \nonumber
&& f(\tau)=\int \frac{d\omega}{2\pi}
\frac{1-e^{-\sqrt{\omega^2+\Delta^2}/T_K}}{\sqrt{\omega^2+\Delta^2}}
\cos(\omega\tau).
\end{eqnarray}
 The $T=0$ supercurrents $I_j$ then come from
 the second-order contribution to the free energy,
 $F^{(2)}=- \frac12\int d\tau \langle {\cal T} H'(\tau) H'(0)\rangle$. Using
 Eq.~\eqref{correlationfunc} and Wick's theorem \cite{SM1}, the phase derivatives of
 $F^{(2)}$ yield
\begin{eqnarray}\label{finalcur}
&& I_j (\varphi_1,\varphi_2,\varphi_3)= I_0\sum_{k\ne j}^3 \Biggl( \sin(\varphi_j-\varphi_k) \\  \nonumber
 && + \frac{1}{3}\sin \left(\frac{\varphi_j+\varphi_k-2\varphi_p}{3} \right)
 - \frac{1}{3}\sin \left(\frac{\varphi_k+\varphi_p-2\varphi_j}{3} \right)\Biggr),
\end{eqnarray}
where $p\ne (j,k)$. The current scale, and thus ultimately the critical current $I_c$, is set by
\begin{equation}
I_0 = \zeta  \frac{\Delta}{T_K} \frac{e\Delta}{\hbar},\quad
 \zeta=\frac{\Lambda_-(\Lambda_- -2\pi)}{3} f^3(0).
\end{equation}
The dimensionless number $\zeta$ is of order unity and can be positive or negative.
Compared to the conventional Kondo system with critical
  current  $I_c=e\Delta/\hbar$ \cite{Glazman1989}, there is a
  suppression factor $\Delta/T_K\ll 1$ due to
  the residual unscreened spin encoded by $\gamma_0$. Equation \eqref{finalcur} obeys  current conservation, $\sum_j I_j=0$, and predicts a $6\pi$-periodic phase dependence which in turn implies charge fractionalization in units of $e^*=2e/3$ for charge transfer between TS leads.  For finite $\Delta$, we have a two-channel
  instead of a four-channel Kondo problem, and hence this value of $e^*$ differs from the one for normal leads probed by shot noise \cite{Zazunov2014,Beri2016}.
   The $6\pi$ periodicity is due to the non-Fermi liquid nature of the two-channel Kondo fixed point and can be seen explicitly by putting $(\varphi_1,\varphi_2,\varphi_3)=(\varphi,\varphi,0)$, where Eq.~\eqref{finalcur} gives $I_{1,2}/I_0=\sin\varphi+  [\sin(\varphi/3)+\sin(2\varphi/3)]/3$.   On the other hand, for $(\varphi_1,\varphi_2,\varphi_3)=(\varphi/2,-\varphi/2,0)$, one gets a
$4\pi$ periodicity, $I_{1,2}/I_0=\pm[\sin(\varphi)+2\sin(\varphi/2)]$,
since the third terminal is now basically decoupled ($I_3=0$).  In general, the $6\pi$ periodicity coexists with $2\pi$ and $4\pi$ effects. Finally, we note that
for an observation of the $6\pi$ Josephson effect, one should probe the supercurrent at finite frequencies, cf.~Ref.~\cite{Bocquillon2016}.

\emph{Conclusions.---}We have studied the Josephson effect through a multi-channel Kondo impurity.  This problem could be realized using a Majorana box device with superconducting leads. The different periodicities in the atomic and the strong-coupling limit ($4\pi$ vs $6\pi$ for three leads) could indicate a quantum phase transition at $\Delta\approx T_K$. This point requires a detailed numerical study which can also clarify to what extent the crossover is universal in $\Delta/T_K$.  It would also be interesting to study  topologically trivial $p$-wave superconductors as leads, and to generalize our strong-coupling analysis to $M>3$ where one may encounter
even higher periodicities in the current-phase relation.

\acknowledgments
We thank C. Mora for discussions. We acknowledge funding by the Deutsche Forschungsgemeinschaft (Bonn) with the network CRC TR 183 (project C04), by the
Brazilian CNPq SwB Program and from MEC-UFRN.

\newpage

\begin{appendix}

\setcounter{equation}{0}
\renewcommand{\theequation}{S\arabic{equation}}

\begin{center}
\textbf{\large Supplementary Material: $\bm{6\pi}$ Josephson effect in Majorana box devices}
\end{center}

\begin{center}
We here provide a detailed derivation of the RG equations and further discuss the strong-coupling solution described in the main text.
\end{center}


\subsection{Derivation of RG equations}

In this part, we present a derivation of the RG equations, allowing for anisotropies in
the gap values $\Delta_j$. Our approach holds for arbitrary $\Delta_j\ll D$, where
$D$ denotes an effective bandwidth for the bulk fermions encapsulated by the
boundary fields $\psi_j$. We first derive the one-loop RG equations without fixing a gauge, and then  choose a convenient gauge where the RG equations depend on the
phases $\varphi_j$ only through initial conditions.
Eventually, the Josephson current-phase relation are expressed in terms of gauge-invariant phase differences $\varphi_j - \varphi_k$, where
charge conservation requires that the action
\begin{equation}  \label{S2}
S[\{\psi_j,\gamma_j\}]=S_{\rm leads}+S_{\rm box}+S_{\rm EC}+S_{\rm CAR}
\end{equation}
is invariant under a global (time- and $j$-independent) gauge transformation,
$\psi_j \to e^{i \theta} \psi_j$ and $\varphi_j \to \varphi_j - 2 \theta$,
with arbitrary phase $\theta$.

The leads are described by [cf.~Eqs.~(4) and (5) in the main text]
\begin{eqnarray}  \nonumber
S_{\rm leads} &= & -\frac12 \int_0^\beta d\tau d\tau' \sum_j \bar \Psi_j(\tau)
G^{-1}_j(\tau-\tau') \Psi_j(\tau') ,\\  \label{sleads}
G_j(\omega) & =& -i\ {\rm sgn}(\omega) \sqrt{1+\frac{\Delta_j^2}{\omega^2}} \
\sigma_0+\frac{\Delta_j e^{-i\varphi_j\sigma_z}}{i\omega} \sigma_x ,
\end{eqnarray}
with the boundary Grassmann fields $\psi_j(\tau)$ and $\bar \psi_j(\tau)$ determining
the Nambu spinor $\Psi_j(\tau)=(\psi_j,\bar\psi_j)^T$.
The box action describes the zero-energy Majorana fields, $S_{\rm box}=\frac12\int d\tau \sum_j \gamma_j\partial_\tau\gamma_j$.
The exchange action reads [cf.~Eq.~(2)]
\begin{equation}
S_{\rm EC}= \int d\tau \sum_{j\ne k}^M \lambda_{jk}
\bar \psi_j \psi^{}_k\gamma_k\gamma_j,
\end{equation}
and the crossed Andreev reflection term is given by [cf.~Eq.~(8)]
\begin{equation}\label{carH}
S_{\rm CAR} = \frac12\int\sum_{j\ne k}
\left( \kappa_{jk} \bar\psi_j\bar \psi_k+\kappa_{jk}^*\psi_j\psi_k\right)\gamma_k\gamma_j  .
\end{equation}
 The $\lambda_{jk}$ and $\kappa_{jk}$ matrix elements (with $j\ne k$)  obey the symmetry constraints
 \begin{equation} \label{symmetryconst}
 \lambda_{jk}=\lambda_{kj}^*,\quad
\kappa_{jk}=\kappa_{kj}.
 \end{equation}

We  start from the functional integral representation of the partition function,
$Z=\int {\cal D}(\psi_j,\bar \psi_j,\gamma_j) e^{-S}.$
Following standard steps \cite{Cardy1996}, we split $\psi_j(\tau) = \psi^<_j + \psi_j^>$ into slow ($\psi_j^<$) and fast ($\psi_j^>$) modes. The Majorana fields
corresponding to the box spin degree of freedom represent a slow degree of freedom.
(However, the same RG equations also follow if one splits the Majorana fields
in the same manner as the $\psi_j$.)
The fields $\psi^<_j$ and $\psi_j^>$ have non-zero Fourier components only for $|\omega| < D/b$ and $D/b<|\omega|< D$,
respectively, with a rescaling parameter  $b>1$.
 Correspondingly, the action $S=S_s+S_f+S_{\rm mix}$ separates into slow, $S_s[\psi^<, \gamma]$, and fast,
$S_f[\psi^>]$, pieces, plus a term $S_{\rm mix}[\psi^>, \psi^<, \gamma]$
which mixes both types of modes.
Next we  integrate out the fast modes,  followed by renaming $\psi_j^<\to \psi_j$ and  rescaling $\omega \to\omega / b$  such that $D$ stays invariant.
Using a cumulant expansion, the renormalized effective action for the slow modes
 is
$\tilde S = S_s[\psi, \gamma] - {1 \over 2} \langle S_{\rm mix}^2 \rangle_>$,
where $\langle \ldots \rangle_>$ denotes the average over the fast modes
and we used $\langle S_{\rm mix} \rangle_> = 0$.  With
$S_{\rm mix}= S_{\rm mix, EC}+S_{\rm mix, CAR}$, by
averaging over the fast modes,  we obtain
\begin{eqnarray}\nonumber
&& \left\langle S_{\rm mix, EC}^{2} \right \rangle_>
=  - \int d \tau \sum_{j \neq n \neq k}
\gamma_k \gamma_j \Bigl(
2 \lambda_{jn} \lambda_{nk} X_n \bar \psi_j \psi_k
\\  && \quad + \label{smix2}
\lambda_{jn} \lambda_{nk}^\ast Y_n \bar \psi_j \bar \psi_k +
\lambda_{jn}^\ast \lambda_{nk} Y_n^\ast \psi_j \psi_k
\Bigr) ,\\ \nonumber
&& \left\langle S_{\rm mix, CAR}^{2}\right \rangle_>
=- \int d \tau \sum_{j \neq n \neq k} \gamma_k \gamma_j \Bigl(
2 \kappa_{jn} \kappa_{nk}^\ast X_n \bar \psi_j \psi_k  \\ \nonumber
&& \quad +
\kappa_{jn} \kappa_{nk} Y_n^\ast \bar \psi_j \bar \psi_k +
\kappa_{jn}^\ast \kappa_{nk}^\ast Y_n \psi_j \psi_k
\Bigr) ,\\ \nonumber
&& \left\langle S_{\rm mix, EC} S_{\rm mix, CAR}\right\rangle_>
= - \int d \tau \sum_{j \neq n \neq k} \gamma_k \gamma_j \ \times \\
\nonumber && \quad \times \Bigl[
\left ( \kappa_{jn} \lambda_{nk} Y_n^\ast  +
 \lambda_{jn} \kappa_{nk}^\ast Y_n \right) \bar \psi_j \psi_k \\ \nonumber
 && \qquad + \
\lambda_{jn} \kappa_{nk} X_n \bar \psi_j \bar \psi_k +
\kappa_{jn}^\ast \lambda_{nk} X_n \psi_j \psi_k
\Bigr] .
\end{eqnarray}
Here we have defined the quantities
\begin{eqnarray}\label{Xj}
X_j &=& \int d \tau \, {\rm sgn}(\tau) \left[ G_j^>(-\tau) \right]_{11} ,\\
\nonumber
Y_j &=& \int d \tau \, {\rm sgn}(\tau) \left[ G_j^>(-\tau) \right]_{12} ,
\end{eqnarray}
where $\left[ G_j^>(-\tau) \right]_{ab}$ is the ($ab$) Nambu matrix component
of the Green function $G_j^>$ for the fast modes. The latter has
the same Fourier components as in Eq.~\eqref{sleads} but is restricted to
the frequency shell $D/b<|\omega| <D$. The
 sign function in Eq.~\eqref{Xj} arises from the time ordering of Majorana operators,
$T_\tau \gamma_j(\tau) \gamma_j(\tau') \to \frac12{\rm sgn}(\tau - \tau')$
for $| \tau - \tau'| \alt 1/D$.
Using $\int d \tau e^{i \omega \tau} {\rm sgn}(\tau) = 2 i / \omega$ for $\omega \neq 0$, we obtain
\begin{equation}\label{xj2}
X_j \simeq \frac{2}{\pi} \sqrt{1 + \delta_j^2} \, \ln b ,\quad
Y_j \simeq \frac{2}{\pi} \delta_j e^{-i \varphi_j} \, \ln b ,
\end{equation}
where we define the real-valued dimensionless gap parameters $\delta_j = \Delta_j / D$.

The rescaling step is completed by renormalizing the scaling variables of the theory.
From Eqs.~\eqref{smix2} and \eqref{xj2},
the renormalized couplings $\lambda_{jk} \to \lambda_{jk} + \delta \lambda_{jk}$
and $\kappa_{jk} \to \kappa_{jk} + \delta \kappa_{jk}$
acquire the running coupling corrections
\begin{eqnarray}
\delta \lambda_{jk} &=& \frac{\ln b}{\pi}
\left( \lambda \rho \lambda + \kappa \rho \kappa^\ast + \kappa
 \tilde \delta^\ast \lambda +  \lambda \tilde \delta \kappa^\ast  +
  {\rm h.c.} \right)_{jk} ,\nonumber \\
\delta \kappa_{jk} &=& \frac{\ln b}{ \pi} \label{runn}
\left( \lambda \tilde \delta \lambda^\ast + \kappa \tilde \delta^\ast \kappa
 + 2 \lambda \rho \kappa \right)_{jk} + (j \leftrightarrow k) ,
\end{eqnarray}
which evidently satisfy the symmetry constraints in Eq.~\eqref{symmetryconst}.
Here, the matrices  $\lambda = \lambda^\dagger$ and $\kappa = \kappa^T$ are
$M \times M$ matrices in lead space with vanishing diagonal elements.
In addition, we have used the diagonal matrices $\tilde \delta$ and $\rho$ with
\begin{equation}
\tilde \delta_{jj} =  e^{-i \varphi_j}  \delta_j , \quad
\quad \rho_{jj} = \sqrt{1 + \delta_j^2}.
\end{equation}
Rescaling $\omega \to\omega/b$ in $S_{\rm leads}$ gives
 $\delta_j\to b \delta_j$, resulting in the scaling equation $d\delta_j/d\ell=\delta_j$ with initial value $\delta_j(0)=\Delta_j/D$ and
the flow parameter $\ell = \ln b$.  Recalling that $D$ is invariant, the
solution is given by
\begin{equation}\label{delflow}
\delta_j(\ell) = e^\ell \frac{\Delta_j}{D}.
\end{equation}
The bulk pairing gaps in the leads thus represent
relevant couplings, which compete with
the Kondo screening processes encoded by $\lambda$ and $\kappa$.
With Eq.~\eqref{runn}, their RG equations are given by
\begin{eqnarray}\label{rgeq1}
\frac{d \lambda}{d \ell} &=& \frac{2}{\pi} \left(
\lambda \rho \lambda + \kappa \rho \kappa^\ast +
\lambda \tilde \delta \kappa^\ast +  \kappa \tilde \delta^\ast \lambda\right) ,\\
  \nonumber
\frac{d \kappa}{ d \ell} &=&\frac{2}{\pi}\left(
 \lambda \tilde \delta \lambda^\ast + \kappa \tilde \delta^\ast \kappa +
\lambda \rho \kappa + \kappa \rho \lambda^\ast\right).
\end{eqnarray}
The equations for $\lambda$ and $\kappa$ are thus coupled through the RG flow of the $\delta_j$.
For $\delta_j = 0$ (normal leads), one recovers the well-known RG
equations for the topological Kondo effect,
$d \lambda / d \ell = (2/\pi) \lambda^2$ with $\kappa
= 0$, where the system flows towards an isotropic strong-coupling fixed point.

We next observe that the RG equations \eqref{rgeq1} are invariant under
local ($j$-dependent) gauge transformations,
\begin{equation}\label{gauge}
\lambda \to e^{i \theta/2} \lambda e^{-i \theta/2} ,\quad
\kappa \to e^{i \theta/2} \kappa e^{i \theta/2} ,\quad
\tilde \delta \to e^{i \theta} \tilde \delta ,
\end{equation}
with an arbitrary diagonal (in lead space)  matrix
$\theta = {\rm diag} \left\{ \theta_j \right\}$.
It is convenient to choose $\theta_j=\varphi_j$ in order
to remove the superconducting phases $\varphi_j$ from
the bulk couplings $\tilde\delta_j$ in the RG equations \eqref{rgeq1}.
We then arrive at
\begin{eqnarray}\label{rgeq2}
\frac{d \lambda}{d \ell} &=& \frac{2}{\pi} \left(
\lambda \rho \lambda + \kappa \rho \kappa^\ast +
\lambda  \delta_\Delta \kappa^\ast +  \kappa  \delta_\Delta \lambda\right) ,\\
  \nonumber
\frac{d \kappa}{ d \ell} &=&\frac{2}{\pi}\left(
 \lambda \delta_\Delta \lambda^\ast + \kappa  \delta_\Delta \kappa +
\lambda \rho \kappa + \kappa \rho \lambda^\ast\right),
\end{eqnarray}
with $\delta_\Delta={\rm diag}(\delta_j)$ and $\delta_j(\ell)$ in Eq.~\eqref{delflow}.
The RG equations \eqref{rgeq2} have the initial conditions $\kappa_{jk}(0)=0$ and
$\lambda_{jk}(0)=(2t_jt^\ast_k/E_C) e^{i(\varphi_j-\varphi_k)/2}$.
Putting all $\Delta_j=\Delta$, we arrive at Eq.~(9) in the main text.

\subsection{Phase-biased RG solution}

In this part, we show analytically that for $\Delta\ll T_K$ and $M=3$, the gauge-invariant
phases $\theta_{jk}=\frac12(\varphi_j+\varphi_k)-\varphi_0$
with $\varphi_0=(\varphi_1+\varphi_2+\varphi_3)/3$ govern the phase dependence
of the crossed Andreev reflection couplings $\kappa_{jk}(\ell)$ when the RG flow approaches the strong-coupling regime.  Here, $j\ne k$ with $j,k=1,2,3$.
We consider the general case
with possibly different gaps $\Delta_j$
in the TS leads, where we require only that at least one of those gaps is different from zero.
By numerical integration of the RG equations, we then find a flow towards the configurations $\lambda_{jk}\to \lambda(\ell) e^{i(\varphi_j-\varphi_k)/2}$ and $\kappa_{jk}(\ell)\to \kappa(\ell) e^{i\theta_{jk}}$, with
the renormalized scalar amplitudes $\lambda(\ell)$ and $\kappa(\ell)$ in Eq.~(11).

We now determine the phase $\theta_{jk}$ using an analytical argument valid in the regime $\lambda\agt \kappa\gg \lambda\delta$,
which is realized when $\Delta_j\ll T_K$ holds for all TS leads. First, we note that
the saturation condition $d\theta_{jk}/d\ell={\rm Im}(d\ln\kappa_{jk}(\ell)/d\ell)=0$
holds when approaching the strong-coupling regime.  Using the RG equations [see Eq.~(9)], with the index $p\ne (j\ne k)$, this gives the conditions
\begin{eqnarray}\label{phaseeq}
&& \sin\left( \frac{\varphi_j-\varphi_p}{2}+\theta_{pk}-\theta_{jk}\right)
+\\
&& + \nonumber\sin\left( \theta_{jp}- \frac{\varphi_p-\varphi_k}{2}-\theta_{jk}\right)=0.
\end{eqnarray}
These conditions are met by
$\theta_{jp}-\theta_{pk}= \frac{\varphi_j-\varphi_k}{2} + \pi n_p$,
where the integers $n_p$ sum to zero, $\sum_{p=1}^3 n_p=0$.  Without loss of generality, we may put $n_p=0$. Hence
Eq.~\eqref{phaseeq} is satisfied by the solution
\begin{equation}
\theta_{jk}= \frac{\varphi_j+\varphi_k}{2} - \varphi_0.
\end{equation}
Next, the global phase $\varphi_0$ is determined by  a gauge invariance argument. Indeed, the RG equations have gauge-invariant initial conditions and preserve gauge invariance.  To ensure that $\theta_{jk}$ remains unchanged under a global shift of all three phases,  we must have $\varphi_0=\sum_j a_j \varphi_j$ with real-valued coefficients $a_j$ subject to the condition $\sum_j a_j=1$.
When none of the gaps $\Delta_j$ vanishes, the RG solution depends only on the phase differences $\varphi_j-\varphi_k$ upon choosing $a_1=a_2=a_3=1/3$.  In the end, we
arrive  at the center-of-mass phase, $\varphi_0=(\varphi_1+\varphi_2+\varphi_3)/3$,  as stated in the main text.  However, if one of the leads represents a normal conductor, say, $\Delta_3=0$, we have $a_3=0$ and $a_1=a_2=1/2$.

\subsection{Strong-coupling solution}

In this part, we provide details concerning the strong-coupling solution near
the two-channel Kondo fixed point for $\Delta_j\ll T_K$ (where at least one $\Delta_j$  must be finite) and $M=3$.
The partition function can be written in the interaction picture as
\begin{equation}
Z=e^{-\beta F}=Z_0 \left\langle {\cal T} e^{-\int_0^\beta d\tau H'(\tau)}\right\rangle,
\end{equation}
where the free part $Z_0=e^{-\beta F_0}$ is defined by the leads alone, i.e., without the
interaction term
\begin{eqnarray}\nonumber
H' &=& H_{\rm LIO}+H_-= 2\pi T_K^{-1/2} \gamma_0 \eta_1(0)\eta_2(0)\eta_3(0)
\\ \label{hprimedef}
&+& \frac{\Lambda_-}{2} T_K^{-1/2} \gamma_0  {\bm\eta}(0) \cdot
[{\bm \xi}(0)\times {\bm \xi}(0) ].
\end{eqnarray}
The perturbation $H'$ is RG-irrelevant with scaling dimension $d=3/2$.  To zeroth order,
the leads are completely decoupled and gauge invariance ensures that the currents $I_j$ do vanish.   The respective Majorana correlation functions for the leads are given by [see Eq.~(14) for $\Delta_j=\Delta$]
\begin{eqnarray}\label{corr1}
&&\langle {\cal T}\eta_j(\tau)\xi_k(0)\rangle = -i
\delta_{jk} \Delta_j\cos(\varphi_j-\varphi_0) f_j(\tau),\\
 \nonumber
&& \langle {\cal T}\eta_j(\tau)\eta_k(0)\rangle =
 \langle {\cal T}\xi_j(\tau)\xi_k(0)\rangle = - \delta_{jk}  \partial_\tau f_j(\tau) ,
\\ \nonumber
&& f_j(\tau)=\int \frac{d\omega}{2\pi}
\frac{1-e^{-\sqrt{\omega^2+\Delta_j^2}/T_K}}{\sqrt{\omega^2+\Delta_j^2}}
\cos(\omega\tau).
\end{eqnarray}
Since we have to contract $\gamma_0$ operators, the first non-vanishing contribution to the free energy arises at second order in $H'$. Furthermore, $\eta$ contractions by themselves do not generate a phase dependence, see Eq.~\eqref{corr1}, and
 we therefore omit the $\langle {\cal T} H_{\rm LIO} (\tau) H_{\rm LIO}(0)\rangle$ contribution below. We then arrive at the perturbation expansion
\begin{eqnarray}\nonumber
F^{(2)} &=& - \int d\tau \langle {\cal T} H_{\rm LIO}(\tau) H_-(0) \rangle \\
&-& \nonumber
\frac12 \int d\tau \langle {\cal T}H_-(\tau) H_-(0)\rangle \\ \label{freeen}
&=&-\sum_{j<k} A_{jk}\cos(\varphi_j-\varphi_0)\cos(\varphi_k-\varphi_0),
\end{eqnarray}
where the coefficients $A_{jk}$, with $p\ne (j,k)$, are given by
\begin{equation}
A_{jk}=- \Lambda_-(\Lambda_--2\pi) \frac{ \Delta_j\Delta_k }{T_K} \int_0^\beta
d\tau f_j(\tau) f_k(\tau) \partial_\tau f_p(\tau).
\end{equation}
Putting all $\Delta_j=\Delta$ and taking the phase derivatives of $F^{(2)}$, we arrive at the current-phase relation in Eq.~(15) with the current scale $I_0$ in Eq.~(16).

Finally, let us briefly discuss the case $\Delta_3=0$, where the lead $j=3$ represents a normal conductor and $I_3=0$ in equilibrium. Putting $\Delta_1=\Delta_2=\Delta$,
the Josephson current flowing in the two TS leads then follows from Eq.~\eqref{freeen}
as
\begin{equation}
I_1=-I_2 = A_{12} \sin(\varphi_1-\varphi_2).
\end{equation}
The $2\pi$ periodicity arises since the normal lead can induce transitions between different parity sectors, in contrast to a standard $4\pi$-periodic topological Josephson junction.

\end{appendix}

\end{document}